\documentclass[a4paper,12pt]{article}

\usepackage{amsmath,amssymb,graphicx}
\usepackage{pst-all}
\usepackage{cite}

\numberwithin{equation}{section}

\newcommand{\dd}{\mathrm{d}}
\newcommand{\pd}{\partial}

\newcommand{\D}{\mathcal{D}}

\newcommand{\e}{\mathrm{e}}
\newcommand{\ket}[1]{\left|#1\right\rangle}

\newcommand{\bracket}[2]{\left\langle%
#1\left.\right|#2\right\rangle}

\newcommand{\tr}{\mathop{\mathrm{tr}}\nolimits}

\newcommand{\I}{\mathbb{I}}
\newcommand{\V}{\mathcal{V}}
\newcommand{\W}{\mathcal{W}}

\newcommand{\Op}{\mathcal{O}}

\newcommand{\ft}[2]{{\textstyle\frac{#1}{#2}}}

\newcommand{\n}{\mathbf{n}}
\newcommand{\m}{\mathbf{m}}
\newcommand{\rr}{\mathbf{r}}
\newcommand{\s}{\mathbf{s}}
\newcommand{\kk}{\mathbf{k}}
\newcommand{\lb}{\mathbf{l}}
\newcommand{\tb}{\mathbf{t}}
\newcommand{\uu}{\mathbf{u}}
\begin{document}

\title{On dilatation operator for a renormalizable theory}%
\author{Corneliu Sochichiu\\
 {\it INFN-Laboratori Nazionali di Frascati}\\
 {\it via Enrico Fermi, 40, Frascati (RM), ITALY}\\
 and\\
 {\it Institutul de Fizic\u{a} Aplicat\u{a}}\\
 {\it str.Academiei, 5,
 MD-2028 Chi\c{s}in\u{a}u, MOLDOVA}\\
e-mail: \texttt{sochichi@lnf.infn.it}
}%


\maketitle
\begin{abstract}
 Given a renormalizable theory we construct the dilatation operator, in the sense of generator of RG flow of composite operators. The generator is found as a differential operator acting on the space of normal symbols of composite operators in the theory. In the spirit of AdS/CFT correspondence, this operator is interpreted as the Hamiltonian of the dual theory. In the case of a field theory with non-abelian gauge symmetry the resulting system is a matrix model.

 The one-loop case is analyzed in details and it is shown that we reproduce known results from $\mathcal{N}=4$ supersymmetric Yang--Mills theory.
\end{abstract}

\tableofcontents
\section{Introduction}

AdS/CFT correspondence \cite{Maldacena:1997re} introduces a correspondence between gauge and string theories. The central role in this correspondence is played by the scale dependence on the gauge side. Namely, under this conjecture, gauge invariant composite operators correspond to physical states of interacting string theory, while their dilatation flow corresponds to the dynamics of respective states. There were numerous checks of the above conjecture, however no complete proof of the conjecture was so far obtained (see \cite{Aharony:1999ti} for a classical review on AdS/CFT correspondence as well as \cite{Maldacena2005,Zaffaroni:2005ty,%
Furuuchi2007,Plefka2006} for updates).

On the other hand, independently of whether AdS/CFT correspondence in its original formulation is true or not, description of scaling properties of quantum field model can be figured out in terms of a dynamical system. In other words one can define a dynamical system whose states are given by composite operators of the field theory model and dynamical evolution is induced by scale transformations. The main point is the form of organization of this dynamics.

In this context huge progress was reached in the analysis of planar $\mathcal{N}=4$ super Yang--Mills theory (see \cite{Beisert:2005} for a review). Thus the dilatation operator was constructed for this theory at one-loop order as a second order differential operator acting on the space of normal symbols of local gauge invariant composite operators as well as a higher order differential operator for some sectors of the theory to higher loops \cite{Beisert:2002bb,Beisert:2002ff,Beisert:2003jj,Beisert:2003tq}.
The planar limit of this operator was found to correspond to integrable spin chain models \cite{Minahan:2002ve,Beisert2003,Beisert:2003tq}, which allows one to make assumptions about the all-order $S$-matrix of the theory \cite{Staudacher:2005}.

Non-planar description can also be given in terms of spin models which generalize the spin chains by inclusion of a chain fusion and fission interaction \cite{Beisert:2002ff,Bellucci:2004ru,Bellucci:2004qx,%
Peeters:2004pt,Bellucci:2004dv}. This interaction is known to break integrability which prevents one from the use of the power of Bethe Ansatz. On the other hand, the knowledge of the dilatation operator in terms of a differential operator on the space of normal symbols of composite operators is enough to have a description in terms of a matrix model and in case when non-planarity plays an important role this description appears to be more natural than one in terms of spin systems \cite{Bellucci:2004fh,Sochichiu:2006uz,Sochichiu:2006yv}.

The progress reached for the $\mathcal{N}=4$ SYM theory would be nice to extend to other cases of gauge/string correspondence e.g. quiver theories \cite{Benvenuti:2004dy} or deformed SYM \cite{Lunin:2005jy}. For the last case a considerable progress was achieved in the study of the scalar sector in the planar limit \cite{Frolov:2005dj,Frolov:2005ty,Frolov:2005iq}. This case has a special significance, since it deals with conformally invariant theories.

In fact, there is an infinite family of possible gauge theories of interest \cite{Lin:2004nb} and it would be useful to have a generic formula or ar at least a simple algorithm allowing us to just plug in the action of the model of interest in order to get the corresponding dilatation operator. Building such an operator is the aim of present study.

We start with broad assumptions about the model: such as free field propagators and superficial renormalizability of the interaction and build the dilatation operator by reducing the perturbative series to few patterns which we call scaling factors. Then the scaling factors are explicitly evaluated. We use the differential renormalization scheme \cite{Freedman:1991tk} in which computation of scaling factors is rather algebraic.

The plan of the paper is as follows. In the next section we give the setup of the problem. Namely, we describe the assumptions we have about of the model, the quantities to be calculated and approach to be used. In the third section we consider the general case of perturbative expansion of dilatation operator in terms of scaling factors. In the fourth section we analyze the one-loop order of the expansion in details computing all scaling factors and giving explicit form of dilatation operator. And finally we discuss our result in the discussion section. Appendices contain useful relations and properties for distribution as well as some technical parts in order to ease the reading of the main body of the text.

\section{The Setup}
Consider the quantum field theory with no dimensional parameter. Assume, that the fundamental excitations of the theory can be parameterized in terms of ``letters'' $\Phi_A$ associated to a space-time point $x$, e.g. $x=0$. The set of letters which contains all elementary fields of the theory as well as  all their derivatives at $x$ we call alphabet. Elementary letters are the fundamental bosonic and fermionic fields having mass dimension 1 and 3/2 respectively. For the fundamental letters we will use the notations $\phi_a$ for bosons and $\psi_{\alpha i}$ for the fermions. The free correlators are,
\begin{equation}
  D_{ab}(x-y)=\rnode{C}{\phi}_a(x)\rnode{D}{\phi}_b(y)
  \ncbar[linewidth=.01,nodesep=2pt,arm=.1,angle=90]{-}{C}{D}=
  \frac{1}{4\pi^2}\frac{\delta_{ab}}{(x-y)^2}.
\end{equation}
for two fundamental bosons as well as
\begin{equation}\label{corr-n}
  D_{(\alpha i)(\beta j)}=
  \rnode{C}{\psi}_{\alpha i}(x)\rnode{D}{\bar{\psi}}_{\beta j}(y)
  \ncbar[linewidth=.01,nodesep=2pt,arm=.1,angle=90]{-}{C}{D}=
  \frac{1}{4\pi^2}(\gamma^{\mu})_{\alpha \beta}\pd^x_\mu\frac{\delta_{ij}}{(x-y)^2},
\end{equation}
for fermions.
Each fundamental letter gives rise to an infinite tower of derivative letters, which we will denote as $\phi_a^{(\n)}$ in the bosonic case and $\psi_{\alpha i}^{(\n)}$ in the fermionic one hiding the Lorenz indices of all derivatives into the superscript $(n)$:
\begin{equation}\label{letter-n}
  \phi_a^{(\n)}\to \pd_{(\mu_1}\dots\pd_{\mu_n)}\phi_a.
\end{equation}
In general throughout this paper a bold-face Latin letters like, $\n,\m,\rr$ etc will mean sets of respectively $n,m,r$ indices. We will treat them as usual sets: $\n+\m$ denotes the union of both sets, $\n-\kk$, where $\kk\in \n$ denotes completion of $\kk$ in $\n$. Since the trace parts of derivatives of the letters corresponding to dynamical variables can be removed by the equations of motion we assume that all derivative letters are traceless. This is denoted by parentheses encircling the indices: $(\n)$.

The correlators of derivative letters are given by acting with respective derivatives on the fundamental correlators, e.g.
\begin{equation}
  D^{(\n)(\m)}_{ab}(x-y)=
  \rnode{C}{\phi}^{(\n)}_a(x)\rnode{D}{\phi}^{(\m)}_b(y)
  \ncbar[linewidth=.01,nodesep=2pt,arm=.1,angle=90]{-}{C}{D}=
  \frac{(-1)^m}{4\pi^2}\pd^{(\n)+(\m)}\frac{\delta_{ab}}{(x-y)^2}.
\end{equation}

Basic objects of our analysis are the composite operators which are products of letters. In the case of gauge systems these operators should also be gauge invariant. If the fundamental degrees of freedom are described by fields in adjoint representation of the gauge group, then the local gauge invariant operators are given by the product of the gauge invariant ``words'' which are traces of product of local fields
\begin{equation}\label{op:def}
  \Op_{A_1A_2\dots A_L}=\tr \Phi_{A_1}\Phi_{A_2}\dots\Phi_{A_L}.
\end{equation}
Otherwise, the composite operators are just polynomials in the fundamental fields.

In the spirit of \cite{Sochichiu:2006uz,Sochichiu:2007eu} we treat the space of composite operators as the Hilbert space of a quantum mechanical system.

This system is further defined by the following data: The rising operator inserting a letter $\Phi_A$, which we (by abuse of notations) call also $\Phi_A$; Lowering operator $\check{\Phi}_A$, which removes a letter from the word; The vacuum state $\ket{\Omega}$, which is annihilated by all lowering operators,
\begin{equation}
  \check{\Phi}_A\ket{\Omega}=0,\qquad \forall \check{\Phi}_A.
\end{equation}

Then, an arbitrary word can be identified with the result of action on the vacuum state of a set of rising operators.

The above definition of the Hermitian product makes letters $\Phi_A$ and $\check{\Phi}_A$ conjugate and, respectively the operator $\Delta_0$ self-conjugate. Another property of the product is that for the field-derivatives-free words it is proportional to the \emph{free} vev of the product of normal ordered operators stripped of $x$-dependence,
\begin{equation}
  \langle :(\Op')^{\dag}::\Op:\rangle_{(0)}=
  \frac{1}{(4\pi^2)^L}\frac{\bracket{\Op'}{\Op}}{
  (x^2)^{\ft12(\Delta[\Op']+\Delta[\Op])}}.
\end{equation}
Generally, it is not true if any of composite operators $\Op$ or $\Op'$ contain a derivative letter, since the product any letter by this choice of scalar product is orthogonal to its derivatives, while at the same time their correlators are non-vanishing.

In this picture the role of time parameter is played by the log of the renormalization scale.

So, in this work we study the scale dependence of operators built of blocks \eqref{op:def}. Classically, the scaling properties are given by the dimensionality of the composite operator, which at its turn is just the sum of dimensionalities of the factors it is made of, e.g.,
\begin{equation}
  \Delta[\Op_{A_1A_2\dots A_L}]=\sum_{k=1}^{L}\Delta[\Phi_{A_k}].
\end{equation}
Therefore, the classical dimensionality is given by the following first order operator,
\begin{equation}\label{class-dim}
  \Delta_{0}=\sum_{\{\Phi_A\}}\Delta_A \tr\Phi_A\check{\Phi}_A,
\end{equation}
where the check denotes the derivative,
\begin{equation}
  \check{\Phi}_A=\frac{\pd}{\pd\Phi_A}.
\end{equation}
As discussed in \cite{Bellucci:2004fh,Sochichiu:2006uz,Sochichiu:2006yv,%
Sochichiu:2007eu}, the classical dimension operator \eqref{class-dim} corresponds to the Hamiltonian of an oscillator system for which each letter represents an oscillation mode.

At the quantum level, however, the composite operators should be renormalized in order to make their correlators finite. This is achieved e.g. by addition of cut-off scale dependent counter-terms.\footnote{We use the differential renormalization scheme, where no explicit addition of counter-terms is needed.} Thus a renormalized version of a composite operator $\Op_I$ is a linear combination mixing it with another composite operators,
\begin{equation}
  \Op_I^{\rm ren}=Z_I^J(\mu)\Op_J,
\end{equation}
where $Z_I^J(\mu)$ define the \emph{mixing matrix} and depends on the cut-off mass scale $\mu$.

The renormalization modifies the scale dependence of the (renormalized) composite operator. Due to this the classical dimension $\Delta_0$ gets corrected by the \emph{anomalous dilatation operator} which is the following matrix,
\begin{equation}\label{d:anom-part}
  H=Z^{-1}(\mu)\cdot \mu \frac{\pd Z(\mu)}{\pd \mu}.
\end{equation}

For a divergent Green function $G$ throughout this paper we will use square brackets to denote the scale dependence of its renormalized part i.e.,
\begin{equation}
  [G]\equiv \mu\frac{\pd G_{\rm ren}}{\pd\mu}.
\end{equation}

As a regularization tool we will use the real space differential renormalization scheme \cite{Freedman:1991tk}.

\subsection{Operator Product Expansion}

The counter-terms needed to renormalize a composite operator $\Op_I$ can be obtained from the analysis of the correlator of $\Op_I$ with any composite operator $\Op$,
\begin{equation}\label{three-norms}
  \langle :\Op::\Op_I:\rangle=\langle :\Op:\e^{\int :V_{\rm int}:}:\Op_I:\rangle_0,
\end{equation}
where the last expectation value is taken in the free theory. The last expression can be evaluated using the Wick theorem. It is given by all possible pair correlators between fields in $:\Op:$, $:\Op':$ and in factors of interaction exponent $\e^{\int :V:}$ in \eqref{three-norms}. We are interested in counter-terms the divergences appearing from the Wick contractions between interaction exponent and the probe composite operator.

The Wick expansion can be suitably encoded into the so called functional form (see  \cite{Kleinert:1996}),
\begin{equation}\label{wick-kleinert}
  \Op=\e^{\pm\frac{1}{2}\int\dd y_1\dd y_2 \frac{\delta}{\delta \phi_{a}} D_{ab}(y_1-y_2)\frac{\delta}{\delta \phi_b}}:\Op:,
\end{equation}
where $\pm$ stands for either fermions or bosons.

The Wick expansion of the product of two normal ordered operators $ :\Op_x(\Phi):$ and $:\Op'_y(\Phi):$ can be written as,
\begin{equation}\label{2-star}
  :\Op_x(\Phi)::\Op'_y(\Phi):=\Op_x*\Op'_y(\Phi),
\end{equation}
where the star-product is given by,
\begin{equation}
  \Op_x(\Phi)*\Op'_y(\Phi)=
  \e^{\check{\Phi}_{Ax}D_{AB}(x-y)\check{\Phi}_{By}}
  :\Op_x(\Phi)\Op'_y(\Phi):,
\end{equation}
where $\check{\Phi}_{xA}$ acts only on $\Op_x$, while $\check{\Phi}_{yB}$ on $\Op'_y$. Equation \eqref{2-star} can be generalized to a triple product which describes the Wick expansion of a a product of three local operators,
\begin{multline}
  :\Op_x::\Op'_y::\Op''_z:=\\
  \exp\{\check{\Phi}_{Ax}D_{AB}(x-y)\check{\Phi}_{By}
  +\check{\Phi}_{Ax}D_{AC}(x-z)\check{\Phi}_{Cz}\\
  +\check{\Phi}_{By}D_{BC}(y-z)\check{\Phi}_{Cz}
  \}
  \Op_x*\Op'_y*\Op''_z
  \\
  \equiv\Op_x*\Op'_y*\Op''_z,
\end{multline}
where the checked letter with a subscript containing $x,y,z$ denotes that only the operator in respective point is differentiated. Generalization to the case of four and more factors is straightforward. Although, the notations look similar to the non-commutative star product (see e.g. \cite{Sochichiu:2002jh,Sochichiu:2005ex} for a review) the star product in our case is perfectly commutative.

Generically, the terms having looping lines are ill defined due to the presence of non-integrable divergences at coinciding points of the correlators. These can be regularized and the singularities removed using an the properties of distributions.

From the functional form of Wick expansion we can figure out that the OPE of the product of the probe operator $:\Op:$ with an arbitrary normal operator $:\Op':$ is an action of a linear differential operator:
\begin{multline}\label{wick-diff}
  \widehat{\Op}_x=
  \Op_x+\D_{AB}(x-y)\left(\frac{\pd \Op}{\pd \Phi_A}\right)_x\check{\Phi}_B\\
  +[D_{AC}(x-y)D_{BD}(x-y)]\left(\frac{\pd^2 \Op}{\pd\Phi_A\pd\Phi_B}\right)_x\check{\Phi}_C\check{\Phi}_D
  +\dots
\end{multline}

Since the regularization and subtraction introduce a dependence on a cut-off parameter $\mu$, the renormalized product will depend on the cut-off:
\begin{multline}\label{ope-diff}
  [:\Op_x(\Phi)::\Op_y'(\Phi):]=
  \left\{[D_{AB}(x-y)]\left(\frac{\pd \Op}{\pd \Phi_A}\right)_x\check{\Phi}_B\right.\\
  \left.+[D_{AC}(x-y)D_{BD}(x-y)]\left(\frac{\pd^2 \Op}{\pd\Phi_A\pd\Phi_B}\right)_x\check{\Phi}_C\check{\Phi}_D
  +\dots\right\}:\Op'_y(\Phi):.
\end{multline}

Now let us apply the rule \eqref{ope-diff} to find the anomalous part coming from the OPE of the product of interaction exponent and the probe operator,
\begin{multline}
  \left[\e^{\int :V_{\rm int}:}:\Op_I:\right]=\\
  \int\dd y[:V_y::\Op_I:]+\frac{1}{2!}\int\dd y_1\int\dd y_2[:V_{y_1}::V_{y_2}::\Op_I:]+\dots
\end{multline}
This can be represented as the result of the action of a linear operator which can be symbolically represented as,
\begin{equation}\label{Delta}
  \Delta=\int\dd y [V_{\rm int}(y)*]+
  \frac{1}{2!}\int\dd y_1\int\dd y_2 [V_{\rm int}(y_1)* V_{\rm int}(y_2)*]+\dots,
\end{equation}
where the operator $[Q*]$ is the scale dependent part of Wick expansion of the star-product of $Q$ (which may contain by itself stars) with the probe operator,
\begin{equation}
  [Q*]\cdot \Op=\mu\frac{\pd}{\pd\mu}[Q*\Op]_{\rm reg}.
\end{equation}

The remaining of this paper is devoted to the detailed analysis and explicit construction of the linear operator \eqref{Delta}.
\section{The General Case}

Consider the first two terms of the dilatation operator \eqref{Delta} in general case.

The first term of \eqref{Delta} is further expanded as,
\begin{multline}\label{D:V}
  \int\dd y\, [V_{\rm int}(y)*]
  =\int\dd y\, \left[\e^{\check{\Phi}_y\cdot D_{y}\cdot\check{\Phi}}\right]V_y\\
  =\int\dd y\left(
  \frac{1}{2}(\check{\Phi}\otimes\check{\Phi})_y\cdot
  [D_{y}\otimes D_{y}]\cdot (\check{\Phi}\otimes\check{\Phi})\right.\\
  +\left.\frac{1}{3!}(\check{\Phi}^{\otimes 3})\cdot [D_{y}^{\otimes 3}]\cdot (\check{\Phi}^{\otimes 3})
  +\frac{1}{4!}(\check{\Phi}^{\otimes 4})\cdot [D_{y}^{\otimes 4}]\cdot (\check{\Phi}^{\otimes 4})+\dots
  \right)V_y,
\end{multline}
where to further shorten the notations we introduced the following notational conventions,
\begin{align}
  \check{\Phi}_y\cdot D_{y-x}\cdot\check{\Phi}_x&=\check{\Phi}_A(y) D_{A B}(y-x)\check{\Phi}(x),\\
  (\check{\Phi}\otimes\check{\Phi})\cdot D\otimes D_{y-x}\cdot(\check{\Phi}\otimes\check{\Phi})&=
  \check{\Phi}_{A_1}\check{\Phi}_{A_2}D_{A_1B_1}D_{A_2B_2}
  \check{\Phi}_{B_1}\check{\Phi}_{B_2},\\
  \Phi^{\otimes n}&=\underbrace{\Phi\otimes\Phi\dots\otimes\Phi}_{n-\text{times}}.
\end{align}
The subscript $y$ in $\check{\Phi}_y$ denotes that respective checked letter acts on the operator in $y$ (in this case $V_y$). At the same time, no subscript means that the letter is localized at $x=0$. Subscript below $D_{y}$ denotes the argument of $D_{AB}(y)$.

In \eqref{D:V} we dropped the linear in $\check{\Phi}$ term, which corresponds to tree level contribution and requires no scale dependent renormalization.

Let us turn now to the two vertex level for which the dilatation operator is given by \eqref{Delta},
\begin{multline}\label{D:VV}
  \frac{1}{2!}\int\dd y_1\int\dd y_2 [V_{\rm int}(y_1)* V_{\rm int}(y_2)*]\\
  =\frac{1}{2!}\int\dd y_1\int\dd y_2
  \left[\e^{\check{\Phi}_{y_1}\cdot D_{y_1-y_2}\cdot\check{\Phi}_{y_2}+\check{\Phi}_{y_1}\cdot D_{y_1}\cdot\check{\Phi}+\check{\Phi}_{y_2}\cdot D_{y_2}\cdot \check{\Phi}}\right]V_{y_1}V_{y_2}.
\end{multline}

As in the one-vertex case not all terms in the expansion of \eqref{D:VV} are relevant for the anomalous dilatation operator. In addition to tree contribution excluded at one point function level there we should exclude also one particle reducible contribution, which should be already taken into consideration by the action and the renormalization at the two point level. Beyond that also the terms corresponding to diagrams containing loops involving only one of two interaction vertices or not involving the probe composite operator should be excluded too, since the counterterms for these diagrams are already taken into account for action renormalization and for one-vertex renormalization.

It is not difficult to check that there are no relevant terms left at the first and second orders of expansion of \eqref{D:VV}. Most terms go away at the third and fourth orders too. The remaining terms at these orders are,
\begin{multline}\label{2nd}
  \frac{1}{2!}\int\dd y_1\int\dd y_2 [V_{\rm int}(y_1)* V_{\rm int}(y_2)*]\\
  =\frac{1}{2}\int\dd y_1\int\dd y_2\times\\
  \biggl\{(\check{\Phi}_{y_1}\otimes\check{\Phi}_{y_1}\otimes
  \check{\Phi}_{y_2})\cdot [D_{y_1}\otimes D_{y_1-y_2}\otimes D_{y_2}]\cdot(\check{\Phi}\otimes \check{\Phi}_{y_2}\otimes\check{\Phi})+\\
  (\check{\Phi}_{y_1}^{\otimes 3}\otimes\check{\Phi}_{y_2})
  \cdot [D_{y_1}^{\otimes 2}\otimes D_{y_1-y_2}\otimes D_{y_2}]\cdot
  (\check{\Phi}^{\otimes 2}\otimes \check{\Phi}_{y_2}\otimes\check{\Phi})
  +\dots\biggr\}V_{y_1}V_{y_2}.
\end{multline}

Thus, to find the generator of dilatations up to the second level, we have to compute the scaling factors of the type $[D_{y_1}D_{y_2}D_{y_1-y_2}]$ for third order and of the type $[D_{y_1}^2D_{y_2}D_{y_1-y_2}]$ for the fourth order. In fact, for fundamental boson  the first term in left hand side of equation \eqref{2nd} is already finite and thus produce no contribution to the dilatation operator, but this is not the case if fermions or derivative letters are involved.

In fact, equations \eqref{D:V} and \eqref{D:VV} already give an idea about the structure of the dilatation operator, while the knowledge of square bracketed parts will fixe the dilatation operator completely.
\section{One-loop order}

Let us restrict ourself to one-loop order in a theory with interaction potential which is of dimension at most four and at most linear in one-derivative letters. This example includes most of the bosonic theories e.g. gauge theories in Feynman-'t Hooft gauge.

The divergent one-loop diagrams which produce non-vanishing contribution to the anomalous part of the dilatation operator appear in the the expansion of the interaction exponential up to second order in interaction potential. In this section we consider first two levels in vertex expansion.

\subsection{One-vertex level}
Application of the one vertex level formula \eqref{D:V} yields,
\begin{multline}\label{comp:V}
  \int\dd y\, [V_{\rm int}(y)*]
  =\int\dd y\, \left[\e^{\check{\Phi}_y\cdot D_{y}\cdot\check{\Phi}}\right]V_y\\
  =\int\dd y\left(
  \frac{1}{2}(\check{\Phi}\otimes\check{\Phi})_y\cdot
  [D_{y}\otimes D_{y}]\cdot (\check{\Phi}\otimes\check{\Phi})\right.\\
  +\left.\frac{1}{3!}(\check{\Phi}^{\otimes 3})\cdot [D_{y}^{\otimes 3}]\cdot (\check{\Phi}^{\otimes 3})
  +\frac{1}{4!}(\check{\Phi}^{\otimes 4})\cdot [D_{y}^{\otimes 4}]\cdot (\check{\Phi}^{\otimes 4})
  \right)V_y,
\end{multline}
where the first term is one-loop, the second and third ones are respectively two and three loop contributions.

Consider the one-loop part of \eqref{D:V} in detail. By the renormalizability $V$ is at most linear in first derivative letters while the composite operators can contain arbitrary number and multiplicity of derivatives. This means that we shall keep in the expansion \eqref{D:V} only the terms that are at most linear in $\check{\phi}^{(1)}_y$, but impose no restrictions on $\check{\phi}^{(n)}$. Thus we have,
\begin{multline}\label{D:V1loop}
  \frac{1}{2}\int\dd y
  (\check{\Phi}\otimes\check{\Phi})_y\cdot
  [D_{y}\otimes D_{y}]\cdot (\check{\Phi}\otimes\check{\Phi})V_y=\\
  \frac{1}{2(4\pi^2)^2}\int \dd y\sum_{\{(\m),(\n)\}}
  (-1)^{n+m}\biggl(
  (\check{\phi}_y\cdot\check{\phi}^{(\n)})
  (\check{\phi}_y\cdot\check{\phi}^{(\m)})
  \left[
  \pd^{(\n)}\frac{1}{y^2}\pd^{(\m)}\frac{1}{y^2}
  \right]+\\
  2 (\check{\phi}^{\mathbf{1}}_y\cdot\check{\phi}^{(\n)})
  (\check{\phi}_y\cdot\check{\phi}^{(\m)})
  \left[
  \pd^{(\n)+\mathbf{1}}\frac{1}{y^2}\pd^{(\m)}\frac{1}{y^2}
  \right]
  \biggr)V_y\equiv\\
  \frac{1}{2(4\pi^2)^2}
  \bigl\{
  \Delta_{(\n),(\m)}(\check{\phi}_a\check{\phi}_b(V))+
  2\Delta_{(\n)+\mathbf{1},(\m)}
  (\check{\phi}^{\mathbf{1}}_a\check{\phi}_b(V))
  \bigl\}\check{\phi}^{(\n)}_a\check{\phi}^{(\m)}_b,
\end{multline}
where we introduced the scaling factors,
\begin{equation}
  \Delta_{\s,\s'}(\V)=
  (-1)^{s+s'}
  \int_x\V_x
  \left[
  \pd^{\s}\frac{1}{x^2}\pd^{\s'}\frac{1}{x^2}
  \right],
\end{equation}
for some main function $\V_x\equiv\V(x)$.

In the second term of the r.h.s of equation \eqref{D:V1loop} we have a derivative $\pd^{(\n)+\mathbf{1}}$, which is not traceless. It can be shown that the trace part of the derivative can be absorbed into a local scale independent counter-term redefinition and therefore does not contribute to the anomalous part of the dilatation operator. Hence, the trace part of $\pd^{(\n)+\mathbf{1}}$ can be safely dropped in \eqref{D:V1loop} replacing $\pd^{(\n)+\mathbf{1}}\to \pd^{(\n+\mathbf{1})}$,
\begin{equation}\label{Delta-rel}
  \Delta_{(\n)+\mathbf{1},(\m)}(\V)=
  -\Delta_{(\n+\mathbf{1}),(\m)}(\V).
\end{equation}

The scaling factor $\Delta_{(\n),(\m)}$ can be evaluated in the following way,
\begin{multline}\label{Delta-nm}
  \Delta_{(\n),(\m)}(\V)=(-1)^{m+n}\int_x\V_x
  \left[
  \pd^{(\n)}\frac{1}{x^2}\pd^{(\m)}\frac{1}{x^2}
  \right]=\\
  2^{n+m}n!m!\int_x \V_x
  \left[\frac{x^{(\n)+(\m)}}{x^{2(n+m+2)}}
  \right]=\\
  2^{n+m}n!m!\sum_{\substack{\rr|\n \\ \rr'|\m}}
  g^{\rr,\rr'}\gamma^{\n\m}_{(\n+\m-2\rr)}\int_x \V_x
  \left[\frac{x^{(\n+\m-2\rr)}}{x^{2(n+m-r+2)}}
  \right]=\\
  -\sum_{\substack{\rr|\n \\ \rr'|\m}}
  \frac{n!m!}{2^{m+n-2r+2}(m+n-2r+1)!(m+n-2r)!}
  g^{\rr,\rr'}\gamma^{\n\m}_{(\n+\m-2\rr)}\\
  \times\int_x \V_x
  \left[x^{(\n+\m-2\rr)}\Box^{m+n-2r+1}\frac{\ln\mu^2x^2}{x^2}
  \right],
\end{multline}
where we used the expansion \eqref{dtr} of product of two traceless representations into the irreducible traceless part and traces.

The scaling factor can be further evaluated to be,
\begin{multline}
  \Delta_{(\n),(\m)}(\V)=\\
  \sum_{\substack{\rr|\n \\ \rr'|\m}}
  \frac{\pi^2n!m!}{2^{m+n-2r-1}(m+n-2r+1)!(m+n-2r)!}
  g^{\rr,\rr'}\gamma^{\n\m}_{(\n+\m-2\rr)}\\
  \times\int_x \V_x
  x^{(\n+\m-2\rr)}\Box^{m+n-2r}\delta(x)\\
  =\sum_{\substack{\rr|\n \\ \rr'|\m}}
  \frac{2\pi^2n!m!}{(m+n-2r+1)!}\
  g^{\rr,\rr'}\gamma^{\n\m}_{(\n+\m-2\rr)}
  \pd^{(\n+\m-2\rr)}\V
  .
\end{multline}

This completes the computation of the one-vertex contribution.

Before closing this section let us note that the the system considerably simplifies if there are no derivative letters in both composite operator and interaction vertex.
In this case we can easily evaluate
all one-vertex scaling factors $[D_y^{\otimes 2,3,4}]$, which correspond to respectively two- and three loop orders.

Indeed,
\begin{multline}\label{Dk}
  [D_y^{\otimes k}]=\frac{\I^{\otimes k}}{(4\pi^2)^k}\left[
  \frac{1}{y^{2 k}}
  \right]=
  -\mu\frac{\pd}{\pd\mu}\frac{\I^{\otimes k}}{(4\pi^2)^k}C_k
  \Box^{k-1}\frac{\ln\mu^2y^2}{y^2}\\
  =\frac{2\I^{\otimes k}}{(4\pi^2)^{k-1}}C_k
  \Box^{k-2}\delta(y).
\end{multline}

For the two equalities of \eqref{Dk} we used the regularization formula \eqref{many-boxes} and the property \eqref{box-delta1} in the Appendix \ref{app:usefull}. The numerical coefficients $C_k$ are given there in \eqref{c1}.

Plugging the result of \eqref{Dk} into \eqref{comp:V} we get for the first level,
\begin{equation}\label{comp:D}
  \int\dd y\, [V_{\rm int}(y)*]
  =\frac{C_2}{4\pi^2}\check{\delta}^2 V-\frac{C_3}{3(4\pi^2)^2}
  \Box\check{\delta}^3V+
  \frac{C_4}{12(4\pi^2)^3}\Box^2\check{\delta}^4V.
\end{equation}
Here we introduced the operator $\check{\delta}$ defined as,
\begin{equation}
  \check{\delta}V=\check{\phi}_a (V)\check{\phi}_a,
\end{equation}
where the first checked letter $\check{\phi}_a$ acts only on $V$.

Now taking the interaction potential $V$ to be one of the scalar self-interaction in $\mathcal{N}=4$ SYM,
\begin{equation}
  V=\frac{g^2}{4}\tr[\phi_a,\phi_b]^2
\end{equation}
we get for the first term in \eqref{comp:D}
\begin{multline}\label{comp:B}
  \frac{1}{16\pi^2}\check{\delta}^2V=\\
  \frac{1}{16\pi^2}
  \tr\left(:[\phi_a,\phi_b][\check{\phi}_a,\check{\phi}_b]:+
  :[\phi_a,\check{\phi}_b][\check{\phi}_a,\phi_b]:+
  :[\phi_a,\check{\phi}_b][\phi_a,\check{\phi}_b]:\right)\\
  =\frac{1}{8\pi^2}\tr\left(:[\phi_a,\phi_b][\check{\phi}_a,\check{\phi}_b]+
  \ft 12:[\phi_a,\check{\phi}_b][\phi_a,\check{\phi}_b]:\right),
\end{multline}
where the checked letters in the colons never act on the non-checked letters within the same group. Also we used that
\begin{equation}
  :[\phi_a,\check{\phi}_a]:\approx 0,
\end{equation}
when acting on gauge invariant composite operators.

Let us note that \eqref{comp:B} is precisely the one-loop $\mathcal{N}=4$ SYM dilatation operator in the compact SO(6) sector found in \cite{Beisert:2002bb}. The remaining terms in \eqref{comp:D} correspond to two- and three-loop contribution coming from the Feynman diagrams with a single interaction vertex.

\subsection{Two-vertex level}
Let us turn to the two vertex contribution \eqref{D:VV} to the dilatation operator. The relevant terms at this level are
\begin{multline}\label{2nd:4}
  \int\dd y_1\int\dd y_2 \times\\
  (\check{\Phi}_{y_1}\otimes \check{\Phi}_{y_1}\otimes\check{\Phi}_{y_2})\cdot
  [D_{y_1}\otimes D_{y_1-y_2}\otimes D_{y_2}]\cdot
  (\check{\Phi}\otimes \check{\Phi}_{y_2}\otimes \check{\Phi})
  V_{y_1}V_{y_2}.
\end{multline}

The one-loop part may be divergent exclusively due to presence of derivative letters. Among the derivative letters present in the composite operator there can be an exchange by an additional derivative from each vertex. Therefore, taking into account the derivative letters the equation \eqref{2nd:4} can be written as,
\begin{multline}\label{2nd:1-all}
  (\check{\Phi}_{y_1}\otimes \check{\Phi}_{y_1}\otimes\check{\Phi}_{y_2})\cdot
  [D_{y_1}\otimes D_{y_1-y_2}\otimes D_{y_2}]\cdot
  (\check{\Phi}\otimes \check{\Phi}_{y_2}\otimes \check{\Phi})=\\
  \sum_{(\n),(\m)}(-1)^{m+n}\times\\
  \biggl(
  (\check{\phi}_{y_1}\otimes \check{\phi}_{y_1}\otimes\check{\phi}_{y_2})\cdot
  [\pd^{(\n)}D_{y_1}\otimes D_{y_1-y_2}\otimes \pd^{(\m)}D_{y_2}]\cdot
  (\check{\phi}^{(\n)}\otimes \check{\phi}_{y_2}\otimes \check{\phi}^{(\m)})\\
  +
  2(\check{\phi}^{\mathbf{1}}_{y_1}\otimes \check{\phi}_{y_1}\otimes\check{\phi}_{y_2})\cdot
  [\pd^{(\n)+\mathbf{1}}D_{y_1}\otimes D_{y_1-y_2}\otimes \pd^{(\m)}D_{y_2}]\cdot
  (\check{\phi}^{(\n)}\otimes \check{\phi}_{y_2}\otimes \check{\phi}^{(\m)})\\
  +2
  (\check{\phi}_{y_1}\otimes
  \check{\phi}^{\mathbf{1}}_{y_1}\otimes
  \check{\phi}_{y_2})\cdot
  [\pd^{(\n)}D_{y_1}\otimes
  \pd^{\mathbf{1}}D_{y_1-y_2}\otimes
  \pd^{(\m)}D_{y_2}]\cdot
  (\check{\phi}^{(\n)}\otimes
  \check{\phi}_{y_2}\otimes
  \check{\phi}^{(\m)})\\
  +
  (\check{\phi}^{\mathbf{1}}_{y_1}\otimes
  \check{\phi}_{y_1}\otimes
  \check{\phi}^{\mathbf{1}'}_{y_2})\cdot
  [\pd^{(\n)+\mathbf{1}}D_{y_1}\otimes
  D_{y_1-y_2}\otimes
  \pd^{(m)+\mathbf{1}'}D_{y_2}]\cdot
  (\check{\phi}^{(\n)}\otimes
  \check{\phi}_{y_2}\otimes
  \check{\phi}^{(\m)})\\
  -
  2(\check{\phi}^{\mathbf{1}}_{y_1}\otimes
  \check{\phi}_{y_1}\otimes
  \check{\phi}_{y_2})\cdot
  [\pd^{(\n)+\mathbf{1}}D_{y_1}\otimes
  \pd^{\mathbf{1}'}D_{y_1-y_2}\otimes
  \pd^{(\m)}D_{y_2}]\cdot
  (\check{\phi}^{(\n)}\otimes
  \check{\phi}^{\mathbf{1}'}_{y_2}\otimes
  \check{\phi}^{(\m)})\\
  -
  (\check{\phi}_{y_1}\otimes
  \check{\phi}^{1}_{y_1}\otimes
  \check{\phi}_{y_2})\cdot
  [\pd^{(\n)}D_{y_1}\otimes
  \pd^{\mathbf{1}+\mathbf{1}'}D_{y_1-y_2}\otimes
  \pd^{(m)}D_{y_2}]\cdot
  (\check{\phi}^{(\n)}\otimes
  \check{\phi}^{\mathbf{1}'}_{y_2}\otimes
  \check{\phi}^{(\m)})
  \biggr).
\end{multline}
Solving the tensor product structure of \eqref{2nd:1-all}
equation \eqref{2nd:1-all} reduces down to,
\begin{multline}\label{2nd:1-all-red}
  (\check{\Phi}_{y_1}\otimes \check{\Phi}_{y_1}\otimes\check{\Phi}_{y_2})\cdot
  [D_{y_1}\otimes D_{y_1-y_2}\otimes D_{y_2}]\cdot
  (\check{\Phi}\otimes \check{\Phi}_{y_2}\otimes \check{\Phi})=\\
  \frac{1}{(4\pi^2)^3}\sum_{(\n),(\m)}(-1)^{m+n}\times\\
  \biggl\{
  (\check{\phi}_{y_1}\cdot\check{\phi}^{(\n)})
  (\check{\phi}_{y_1}\cdot \check{\phi}_{y_2})
  (\check{\phi}_{y_2}\cdot \check{\phi}^{(\m)})
  \left[\pd^{(\n)}\frac{1}{y_1^2}
  \frac{1}{(y_1-y_2)^2}
  \pd^{(\m)}\frac{1}{y_2^2}\right]
  \\
  +2(\check{\phi}^{\mathbf{1}}_{y_1}\cdot \check{\phi}^{(\n)}) (\check{\phi}_{y_1}\cdot\check{\phi}_{y_2})
  (\check{\phi}_{y_2}\cdot \check{\phi}^{(\m)})
  \left[\pd^{(\n)+\mathbf{1}}\frac{1}{y_1^2}
  \frac{1}{(y_1-y_2)^2}
  \pd^{(\m)}\frac{1}{y_2^2}\right]\\
  +2(\check{\phi}_{y_1}\cdot \check{\phi}^{(\n)})
  (\check{\phi}^{\mathbf{1}}_{y_1}\cdot
  \check{\phi}_{y_2}) (\check{\phi}_{y_2}\cdot
  \check{\phi}^{(\m)})
  \left[\pd^{(\n)}_{\mu}\frac{1}{y_1^2} \pd^{1}\frac{1}{(y_1-y_2)^2}
  \pd^{(\m)}\frac{1}{y_2}\right]
  \\
  +(\check{\phi}^{1}_{y_1}\cdot\check{\phi}^{(\n)}) (\check{\phi}_{y_1}\cdot\check{\phi}_{y_2})
  (\check{\phi}^{\mathbf{1}' }_{y_2}\cdot\check{\phi}^{(\m)})
  \left[\pd^{(\n)+\mathbf{1}}\frac{1}{y_1^2}
  \frac{1}{(y_1-y_2)^2}
  \pd^{(\m)+\mathbf{1}'}_{\nu}\frac{1}{y_2^2}\right]\\
  -2(\check{\phi}^{1}_{y_1}\cdot \check{\phi}^{(\n)}) (\check{\phi}_{y_1}\cdot \check{\phi}^{\mathbf{1}'}_{y_2})
  (\check{\phi}_{y_2}\cdot \check{\phi}^{(\m)})
  \left[\pd^{(\n)+\mathbf{1}}\frac{1}{y_1^2} \pd^{\mathbf{1}'}_{\nu}\frac{1}{(y_1-y_2)^2}
  \pd^{(\m)}\frac{1}{y_2^2}\right]\\
  -(\check{\phi}_{y_1}\cdot\check{\phi}^{(\n)})
  (\check{\phi}^{\mathbf{1}}_{y_1}\cdot \check{\phi}^{\mathbf{1}'}_{y_2})
  (\check{\phi}_{y_2}\cdot \check{\phi}^{(\m)})
  \left[\pd^{(\n)}\frac{1}{y_1^2}
  \pd^{\mathbf{1}+\mathbf{1}'}\frac{1}{(y_1-y_2)^2} \pd^{(\m)}\frac{1}{y_2}\right]
  \biggr\}.
\end{multline}
Let us denote the scaling factors appearing in \eqref{2nd:1-all-red} respectively as $(-1)^{m+n}\Delta_{(\n),0,(\m)}$, $(-1)^{m+n}\Delta_{(\n)+1,0,(\m)}$, $(-1)^{m+n}\Delta_{(\n),1,(\m)}$, $(-1)^{m+n}\Delta_{(\n)+1,0,(\m)+1'}$ etc., depending on the presence of derivatives in propagators.

Consider the first scale factor smeared with two probe functions $\V(x)\equiv \V_x$ and $\W(y)\equiv \W_y$,
\begin{equation}
  \Delta_{(\n),0,(\m)}(\V,\W)=(-1)^{m+n}\int\dd x\dd y\,\V_x\W_y
  \left[\pd^{(\n)}\frac{1}{x^2}
  \frac{1}{(x-y)^2}
  \pd^{(\m)}\frac{1}{y^2}\right],
\end{equation}
and do the following formal manipulations (dropping the divergent divergence terms):
\begin{multline}\label{D:VW}
  \Delta_{(\n),0,(\m)}(\V,\W)=
  \int_{xy} \frac{1}{x^2y^2}
  \pd_x^{(\n)}\pd_y^{(\m)}\left\{
  \V_x\W_y\frac{1}{(x-y)^2}
  \right\}=\\
  \int_{xy} \frac{1}{x^2y^2}
  \sum_{\substack{\kk|\n\\ \lb|\m}}
  \pd_x^{(\n-\kk)}\V_x
  \pd_y^{(\m-\lb)}\W_y
  \pd_x^{(\kk)}\pd_y^{(\lb)}\frac{1}{(x-y)^2}=\\
  \int_{xy} \frac{1}{x^2y^2}
  \sum_{\substack{\kk|\n \\ \lb|\m}}(-1)^l
  \pd_x^{(\n-\kk)}\V_x
  \pd_y^{(\m-\lb)}\W_y
  \pd_x^{(\kk)+(\lb)}\frac{1}{(x-y)^2},
\end{multline}
where in the second equality we used the Leibnitz rule \eqref{Leibnitz-rule} for multiple derivatives. For the last equality of \eqref{D:VW} we used the possibility to trade the $y$-derivatives of $1/(x-y)^2$ for $x$-derivatives at the price of an extra minus factor.

As a product of two traceless representation can be expanded into irreducible traceless representations and traces (see \eqref{trtr-tr}) and this can be applied to derivatives too, we have,
\begin{equation}\label{dtr}
  \pd^{(\n)+(\m)}=\sum_{\substack{\rr|\n\\ \rr'|\m\\
  r=r'}}g^{\rr,\rr'}
  \gamma^{\n\m}_{\n+\m-2 \rr}\pd^{(\n+\m-2\rr)}\Box^{r},
\end{equation}
where the sum runs over two partitions $\rr|\n$ and $\rr'|\m$  of $\n$ and $\m$ respectively, having the same length $r\leq \min\{n,m\}$ and $g^{\rr,\rr'}$ is the product of metric components with first index in $\rr$ and second in $\rr'$.

Plugging \eqref{dtr} into equation \eqref{D:VW} we have,
  \begin{multline}\label{D:VW1}
  \Delta_{(n),0,(m)}(\V,\W)=
  \int_{xy} \frac{1}{x^2y^2}\times\\
  \sum_{\substack{\kk|\n\\ \lb|\m}}(-1)^l
  \sum_{\substack{\rr|\kk\\ \rr'|\lb\\
  r=r'}}g^{\rr,\rr'}\gamma^{\kk\lb}_{\kk+\lb-2\rr}
  \pd_x^{(\n-\kk)}\V_x
  \pd_y^{(\m-\lb)}\W_y
  \pd_x^{(\kk+\lb-2 \rr)}\Box^{r}\frac{1}{(x-y)^2}=\\
  -4\pi^2\int_{xy} \frac{1}{x^2y^2}
  \sum_{\substack{\kk|\n\\ \lb|\m}}(-1)^l
  \sum_{\substack{\rr|\kk\\ \rr'|\lb\\
  r=r'}}
  g^{\rr,\rr'}\gamma^{\kk\lb}_{\kk+\lb-2\rr}
  \pd_x^{(\n-\kk)}\V_x
  \pd_y^{(\m-\lb)}\W_y\times\\
  \pd_x^{(\kk-\rr+\lb-\rr')}
  \Box^{r-1}\delta(x-y)=\\
  -4\pi^2\int_{x}\sum_{\substack{\kk|\n\\ \lb|\m}}(-1)^{k}
  \sum_{\substack{\rr|\kk\\ \rr'|\lb\\
  r=r'}}v\gamma^{\kk\lb}_{\kk+\lb-2\rr}
  \frac{1}{x^2}
  \pd_x^{(\n-\kk)}\V_x
  \pd_x^{(\kk-\rr+\lb-\rr')}
  \Box^{r-1}\left\{\frac{1}{x^2}
  \pd_x^{(\m-\lb)}\W_x
  \right\}
  ,
\end{multline}
where in order to get the last equality we integrated out $y$ using the $\delta$-function.

Let us consider the factor
\begin{equation}
  \pd_x^{(\kk+\lb-2\rr)}
  \Box^{r-1}\left\{\frac{1}{x^2}
  \pd_x^{(\m-\lb)}\W_x
  \right\},
\end{equation}
in the last line of \eqref{D:VW1}. Applying the Leibnitz rule to it, we can redistribute the derivatives among $1/x^2$ and $\W$,
\begin{multline}\label{Leibnitz-box}
  \pd_x^{(\kk+\lb-2 \rr)}
  \Box^{r-1}\left\{\frac{1}{x^2}
  \pd_x^{(\m-\lb)}\W_x
  \right\}=\\
  \pd^{(\kk+\lb-2\rr)}\pd^{\rr-1}
  \sum_{\s|\rr-1}
  \pd^{\s}\frac{1}{x^2}\pd^{\rr-1-\s+(\m-\lb)}\W=\\
  \sum_{\substack{\s|\rr-1\\ \tb|\rr-1 \\ \uu|\kk-\rr+\lb-\rr'}}
  \pd^{\s+\tb+\uu}\frac{1}{x^2}
  \pd^{\rr-1-\s+\rr-1-\tb+(\kk+\lb-2\rr-\uu)+(\m-\lb)}\W.
\end{multline}

The derivatives acting on $1/x^2$ can be expanded in traceless components and traces. It is not very difficult to see that any trace leads to a local subtraction scale independent counter-term and can be discarded. Indeed, trace-full contribution in \eqref{D:VW1} contains a local factor
\begin{equation}
  \sim \frac{1}{x^2}\pd^{(\mathbf{w})}\Box^{p}\delta(x)
  \sim \pd^{(\mathbf{w}')}\Box^{p+1}\delta(x),
\end{equation}
which has no dependence on subtraction scale $\mu$. Hence we can restrict ourself to the analysis of the traceless part of $\pd^{\s+\tb+\uu}$ in \eqref{Leibnitz-box}. Thus, the scale factor reads,
\begin{multline}\label{D:VW2}
  \Delta_{(\n),0,(\m)}(\V,\W)=
  -4\pi^2\int_{x}\sum_{\substack{\kk|\n \\ \lb|\m}}(-1)^{k+l}
  \sum_{\substack{\rr|\kk\\ \rr'|\lb\\
  r=r'}}g^{\rr,\rr'}\gamma^{\kk\lb}_{\kk+\lb-2\rr}
  \sum_{\substack{\s|\rr-1\\ \tb|\rr-1 \\ \uu|\kk-\rr+\lb-\rr'}}\times\\
  \pd^{(\n-\kk)}\V
  \pd^{\rr-1-\s+\rr-1-\tb+(\kk+\lb-2\rr-\uu)+(\m-\lb)}\W
  \left[
  \frac{1}{x^2}
  \pd^{(\s+\tb+\uu)}
  \frac{1}{x^2}
  \right].
\end{multline}
The singularity was isolated in the square brackets and can be evaluated to be,
\begin{multline}\label{VW:sing}
  \left[
  \frac{1}{x^2}
  \pd^{(\s+\tb+\uu)}
  \frac{1}{x^2}
  \right]=(-1)^{s+t+u} 2^{s+t+u}(s+t+u)!\left[
  \frac{x^{(\s+\tb+\uu)}}{x^{2(s+t+u+2)}}
  \right]=\\
  \frac{(-1)^{s+t+u+1}x^{(\s+\tb+\uu)}}{2^{2(s+t+u)+2}(s+t+u)!}
  \left[
  \Box^{s+t+u+1}\frac{\ln\mu^2x^2}{x^2}
  \right]=\\
  2\pi^2\frac{(-1)^{s+t+u}}{s+t+u+1}\pd^{(\s+\tb+\uu)}\delta(x).
\end{multline}

Plugging \eqref{VW:sing} into \eqref{D:VW2} we get,
\begin{multline}\label{D:VW3}
  \Delta_{(\n),0,(\m)}(\V,\W)=\\
  -8\pi^4\sum_{\substack{\kk|\n \\l|\m}}(-1)^{k}
  \sum_{\substack{\rr|\kk\\ \rr'|\lb\\
  r=r'}}g^{\rr,\rr'}\gamma^{\kk\lb}_{\kk+\lb-2\rr}
  \sum_{\substack{\s|\rr-1\\ \tb|\rr-1 \\ \uu|\kk-\rr+\lb-\rr'}}\frac{1}{s+t+u+1}\times\\
  \pd^{(\s+\tb+\uu)}
  \left\{\pd^{(\n-\kk)}\V
  \pd^{\rr-1-\s+\rr-1-\tb+(\kk+\lb-2\rr-\uu)+(\m-\lb)}\W
  \right\}.
\end{multline}

Using \eqref{D:VW3}, we can readily write the first term in the r.h.s. of \eqref{2nd:1-all-red},
\begin{equation}\label{2nd:1st}
  \frac{1}{8\pi^2}
  (\check{\phi}_1\cdot\check{\phi}^{(\n)})
  (\check{\phi}_1\cdot \check{\phi}_2)
  (\check{\phi}_2\cdot \check{\phi}^{(\m)})
  \Delta_{(\n),0,(\m)}(V_1,V_2),
\end{equation}
where the checked letter $\check{\phi}_{i}$, $i=1,2$ acts only on the respective vertex $V_i$ \emph{before} the action of other derivatives i.e. \eqref{2nd:1st} reads,
\begin{equation}
  \frac{1}{8\pi^2}
  \Delta_{(\n),0,(\m)}(\check{\phi}_a\check{\phi}_b ( V),\check{\phi}_b\check{\phi}_c(V))
  \check{\phi}_a^{(\n)}\check{\phi}_c^{(\m)}.
\end{equation}

The remaining scaling factors in \eqref{2nd:1-all-red} are evaluated in a similar way. Here we just state the results,
\begin{subequations}\label{Deltas}
\begin{multline}
  \Delta_{(\n)+\mathbf{1},0,(\m)}(\V,\W)=\\
  -8\pi^4\sum_{\substack{\kk|\n \\ \lb|\m}}(-1)^{k}
  \sum_{\substack{\rr|\kk\\ \rr'|\lb\\
  r=r'}}g^{\rr,\rr'}\gamma^{\kk\lb}_{\kk+\lb-2\rr}\biggl(
  \sum_{\substack{\s|\rr-\mathbf{1}\\ \tb|\rr-\mathbf{1} \\ \uu|\kk+\lb-2\rr}}
  \frac{1}{s+t+u+1}\times\\
  \pd^{(\s+\tb+\uu)}
  \biggl\{\pd^{(\n-\kk)+\mathbf{1}}\V
  \pd^{\rr-\mathbf{1}-\s+\rr-\mathbf{1}-\tb+(\kk+\lb-2\rr-\uu)
  +(\m-\lb)}\W\\
  +\pd^{(\n-\kk)}\V
  \pd^{\rr-\mathbf{1}-\s+\rr-\mathbf{1}-\tb+(\kk+\lb-2 \rr-\uu)+(\m-\lb)+1}\W
  \biggr\}\\
  +\sum_{\substack{\s|\rr-\mathbf{1}\\ \tb|\rr-\mathbf{1} \\ \uu|\kk+\lb-2\rr+\mathbf{1}}}
  \frac{1}{s+t+u+1}\times\\
  \pd^{(\s+\tb+\uu)}
  \biggl\{\pd^{(\n-\kk)+\mathbf{1}}\V
  \pd^{\rr-\mathbf{1}-\s+\rr-\mathbf{1}-\tb+(\kk+\lb-2\rr+
  \mathbf{1}-\uu)+(\m-\lb)}\W
  \biggr\}
  \biggr)=\\
  -\Delta_{(\n),0,(\m)}(\pd^{\mathbf{1}}\V,\W)
  -\Delta_{(\n),\mathbf{1},(\m)}(\V,\W),
\end{multline}
\begin{multline}
  \Delta_{(\n),1,(\m)}(\V,\W)=\\
  8\pi^4\sum_{\substack{\kk|\n \\ \lb|\m}}(-1)^{k}
  \sum_{\substack{\rr|\kk\\ \rr'|\lb\\
  r=r'}}g^{\rr,\rr'}\gamma^{\kk\lb}_{\kk+\lb-2\rr}\biggl(
  \sum_{\substack{\s|\rr-1\\ \tb|\rr-1 \\ \uu|\kk+\lb-2\rr'}}
  \frac{1}{s+t+u+1}\times\\
  \pd^{(\s+\tb+\uu)}
  \biggl\{
  \pd^{(\n-\kk)}\V
  \pd^{\rr-1-\s+\rr-1-\tb+(\kk+\lb-2\rr-\uu)+(\m-\lb)+1}\W
  \biggr\}\\
  +\sum_{\substack{\s|\rr-1\\ \tb|\rr-1 \\ \uu|\kk+\lb-2\rr+1}}
  \frac{1}{s+t+u+1}\times\\
  \pd^{(\s+\tb+\uu)}
  \biggl\{\pd^{(\n-\kk)+1}\V
  \pd^{\rr-1-\s+\rr-1-\tb+(\kk+\lb-2\rr+1-\uu)+(\m-\lb)}\W
  \biggr\}
  \biggr),
\end{multline}
\begin{multline}
  \Delta_{(\n)+\mathbf{1},0,(\m)+\mathbf{1}'}(\V,\W)=
  \Delta_{(\n),0,(\m)}(\pd^{\mathbf{1}}\V,\pd^{\mathbf{1}'}\W)
  +\Delta_{(\n),\mathbf{1},(\m)}(\V,\pd^{\mathbf{1}'}\W)\\
  -\Delta_{(\n),\mathbf{1}',(\m)}(\pd^{\mathbf{1}}\V,\W)
  -\Delta_{(\n),\mathbf{1}+\mathbf{1}',(\m)}(\V,\W).
\end{multline}
\begin{equation}
  \Delta_{(\n)+\mathbf{1},\mathbf{1}',(\m)}(\V,\W)=
  -\Delta_{(\n),\mathbf{1},(\m)}(\pd^{1}\V,\W)
  -\Delta_{(\n),\mathbf{1}+\mathbf{1}',(\m)}(\V,\W),
\end{equation}
\begin{multline}
  \Delta_{(\n),\mathbf{1}+\mathbf{1}',(\m)}(\V,\W)=\\
  -8\pi^4 \sum_{\substack{\kk|\n \\ \lb|\m}}(-1)^{k}
  \sum_{\substack{\rr|\kk,\lb\\ \tilde{\rr}|\kk+\lb-2 \rr+\mathbf{1}+\mathbf{1'}}}
  g^{\rr,\rr'}g^{\tilde{\rr},\tilde{\rr}'}
  \gamma^{\kk,\lb}_{\kk+\lb-2\rr}
  \gamma^{\mathbf{1}+\mathbf{1'},\kk+\lb-2\rr}_{\kk+\lb-2\rr
  +\mathbf{1}+\mathbf{1'}-2\tilde{\rr}}\times\\
  \sum_{\substack{\s,\tb|\rr+\tilde{\rr}-\mathbf{1}\\
  \uu|\kk+\lb-2\rr+\mathbf{1}+\mathbf{1'}-\tilde{\rr}}}
  \frac{\pd^{(\s+\tb+\uu)}}{s+t+u+1}\times\\
  \bigl\{
  \pd^{(\n-\kk)}\V\pd^{\rr+\tilde{\rr}-1-\s+\rr+\tilde{\rr}-1-\tb
  +(\kk+\lb-2\rr+\mathbf{1}+\mathbf{1'}-2\tilde{\rr}-\uu)+(\m-\lb)}\W
  \bigl\}\\
  -2 \pi^4(-1)^{m+n}g^{\mathbf{1},\mathbf{1'}}
  \sum_{\substack{\kk|\n \\ \lb|\m}}(-1)^{l}
  \sum_{\rr|(\kk)+(\lb)}\frac{\pd^{(\rr)}}{r+1}\bigl\{
  \pd^{(\n-\kk)}\V\pd^{(\m)+(\kk)-\rr}\W
  \bigr\}.
\end{multline}
\end{subequations}

Summarizing, we can write down the the two-vertex one-loop part of the dilatation operator in the following form,
\begin{multline}\label{2nd:1-all-fin}
  \int_{y_1y_2}(\check{\Phi}_{y_1}\otimes \check{\Phi}_{y_1}\otimes\check{\Phi}_{y_2})\cdot
  [D_{y_1}\otimes D_{y_1-y_2}\otimes D_{y_2}]\cdot
  (\check{\Phi}\otimes \check{\Phi}_{y_2}\otimes \check{\Phi})V_1V_2=\\
  \frac{1}{(4\pi^2)^3}\sum_{(n),(m)}
  \biggl\{
  \Delta_{(\n),0,(\m)}(\check{\phi}_a\check{\phi}_b(V),
  \check{\phi}_b\check{\phi}_c(V))
  \\
  +2\Delta_{(\n)+\mathbf{1},0,(\m)}(\check{\phi}^{\mathbf{1}}_a
  \check{\phi}_b(V),
  \check{\phi}_b\check{\phi}_c(V))
  +2 \Delta_{(\n),\mathbf{1},(\m)}(\check{\phi}_a
  \check{\phi}^{\mathbf{1}}_b(V),
  \check{\phi}_b\check{\phi}_c(V))
  \\
  +\Delta_{(\n)+\mathbf{1},0,(\m)+\mathbf{1'}}
  (\check{\phi}^{\mathbf{1}}_a\check{\phi}_b(V),
  \check{\phi}_b\check{\phi}^{\mathbf{1'}}_c(V))\\
  -2 \Delta_{(\n)+\mathbf{1},\mathbf{1'},(\m)}
  (\check{\phi}^{\mathbf{1}}_a\check{\phi}_b(V),
  \check{\phi}^{\mathbf{1'}}_b\check{\phi}_c(V))\\
  -\Delta_{(\n),\mathbf{1}+\mathbf{1'},(\m)}
  (\check{\phi}_a\check{\phi}^{\mathbf{1}}_b(V),
  \check{\phi}^{\mathbf{1'}}_b\check{\phi}_c(V))
  \biggr\}\check{\phi}^{(\n)}_a\check{\phi}^{(\m)}_c,
\end{multline}
where various $\Delta_{\s,\s'',\s'''}$ are given in \eqref{D:VW3} and \eqref{Deltas}.

\subsection{Fermionic contribution}

So far we analyzed the case of purely bosonic exchanges. Due to space restriction the fermionic contribution is not discussed here. However, let us give some technical hints.

To include the fermionic contribution coming from both composite operators and interaction vertex we have to do the above computation replacing the bosonic propagator with the fermionic one \eqref{corr-n} and taking into account the signs due to fermion anti-commutative nature.

The scaling factors corresponding to exchange of fermions are basically the same (up to a sign) as those for exchanging a derivative-free letter into a derivative one or viceversa. The difference is that the fermionic checked letters are contracted by a $\gamma$-matrices in contrast to just $\delta$-symbols in the bosonic case.

\subsection{Gauge invariance}
When considering a gauge theory it is important to ensure that the dilatation operator acts within the subspace of gauge invariant composite operators. In what concerns global gauge invariance it is more or less straightforward to see that it is respected by the dilation operator.

The local gauge invariance is more subtle. As we know, it can be broke by quantum corrections, but if there is a explicitly gauge invariant renormalization scheme this is not the case (see e.g. \cite{sf}). As the theory was assumed to be a renormalizable and gauge anomaly free, such a scheme is implied to exist too.

\section{Discussion}
In this work we gave an explicit construction of the generator of RG dilatations for local composite operators in a theory with superficially renormalizable interaction. The one-loop result is given in equations \eqref{D:V1loop}, \eqref{Delta-rel} and \eqref{Delta-nm} for one vertex contribution and
\eqref{2nd:1-all-fin} for the two vertices. For a superficially renormalizable theory there are no expected divergencies and therefore no anomalous contribution to the dilatation operator beyond the two-vertices at one-loop order.

The obtained results can be applied to any renormalizable scalar or gauge theory. The inclusion of fermions is straightforward and does not imply any additional computation.
Although we use condensed notations for Lorenz group\footnote{In the group of $4D$ rotations, since we are are working in the Euclidean signature.} indices which allows us writing heavy expressions in a relatively compact form, inclusion of fermions still complicates considerably the output. We hope that more algebraic approach can be used instead which would allow treat the fermions at the same footing as bosonic fields saving the complexity of expressions.

Another task would be extending present results to the two-loop order and higher. In the present work we analyzed the higher loop contribution coming from a single vertex extension. The higher loop contribution coming from multi-vertex diagrams introduce new technical difficulties related to overlapping divergencies. Of course, there are well developed tools to treat such a problem. One may e.g. use the approach of \cite{Beisert:2003tq} and \cite{Beisert:2007hz} to iteratively subtract the divergences\footnote{I am grateful to the Referee for pointing me to this possibility.} or use a combination of this approach with differential regularization used in this work. 

On the other hand the higher loop diagrams with overlapping divergences produce alongside with the log of scale contribution also the higher powers of the log, which correspond to subtractions in divergent sub-diagrams. By tuning the subgraph subtraction scale one can absorb the linear log term.  It would be interesting to understand wether this implies that the contribution from the diagrams with overlapping divergences reduces to a superposition of contributions of their lower order sub-diagrams. In this case the complete all-loop dilatation operator could be described in terms of a finite number of basic diagrams. Such basic diagrams appear only up to four loop order.

\subsection*{Acknowledgements}
At different stages of this work I benefited from useful discussions with P.-Y.~Casteill, C.~Corian\`o, C.~Kristjansen, Y.~Stanev, M.~Staudacher.

This work is entirely supported by the
European Community's Human Potential Program under contract MRTN-CT-2004-005104 'Constituents, fundamental forces and symmetries of the universe'.
\newpage
\appendix
\section{Useful formulae}\label{app:usefull}
Following formula are useful for differential renormalization.
\begin{align}\label{box-delta1}
  \Box\frac{1}{x^2}&=-4\pi^2\delta(x),\\
  \label{many-boxes}
  \frac{1}{x^{2k}}&=-\frac{1}{4^{k-1}(k-1)!(k-2)!}\Box^{k-1}
  \frac{\ln\mu^2 x^2}{x^2}, \qquad k\geq 2.
\end{align}
In particular,
\begin{equation}
  \frac{1}{x^4}=-\frac{1}{4}\Box\frac{\ln\mu^2 x^2}{x^2}.
\end{equation}
The numerical coefficient in in front of r.h.s. of \eqref{many-boxes} we denote by $C_k$ i.e.,
\begin{equation}\label{c1}
  C_k=\frac{1}{4^{k-1}(k-1)!(k-2)!}.
\end{equation}

Let us also give the expression for the scaling dependence of \eqref{many-boxes},
\begin{equation}\label{many-b-d}
  \left[\frac{1}{x^{2k}}\right]=
  8\pi^2 C_k \Box^{k-2}\delta(x),
\end{equation}
where we used eq. \eqref{box-delta1}.

\subsubsection*{Leibnitz rule}
Often we apply the Leibnitz rule, which in the case of multiple derivatives is spelled out as,
\begin{equation}\label{Leibnitz-rule}
  \pd^{(\n)}(f\cdot g)=\sum_{\kk|\n}\pd^{(\n-\kk)}f\pd^{(\kk)}g,
\end{equation}
where sum spans all ordered partitions $\kk|\n$ of the set of $n$ indices. The sets $\kk$ and $\n-\kk$ are respectively the subsets of taken apart and left indices.

\subsubsection*{Representation decomposition}
Another fact which we use is that the product of two traceless homogeneous polynomials of $x^\mu$ can be expanded in products of traceless polynomials of lower order and traces,
\begin{equation}
  x^{(\mu_1}\dots x^{\mu_n)}
  x^{(\nu_1}\dots x^{\nu_m)}=\sum_{j=0}^{\min(m,n)}
  \gamma^{\mu_1\dots \mu_n,\nu_1\dots\nu_m}_{\alpha_1\dots\alpha_{m+n-2j}}
  (x^2)^jx^{(\alpha_1}\dots x^{\alpha_{m+n-2j})},
\end{equation}
or in the short form,
\begin{equation}\label{trtr-tr}
  x^{(\n)}x^{(\m)}=\sum_{\substack{\rr|\n \\
  \rr'|\m}}
  g^{\rr,\rr'}
  \gamma^{\n \m}_{(\m+\n-2\rr)}
  x^{2j}x^{(\m+\n-2\rr)}.
\end{equation}
Taking into account, that the traceless homogeneous polynomials are when properly normalized in fact 3-spherical harmonics while the trace corresponds to the singlet representation of the four-dimensional rotation group, the coefficients $\gamma^{\n \m}_{\n+\m-2\rr}$ are given by (four dimensional analogues of) Clebsch-Gordan coefficients.

\section{Over-regularization consistency}
When regularizing the singular Green's functions $1/x^{2k}$ one my decide whether to include or not certain polynomials of $x$ inside the regulator factor or to keep it as a outside multiplier. In particular, this polynomial may include various powers of $x^2$ itself which affects the choice of the regularization \eqref{many-boxes}. Let us prove here, that the result for the scale dependence is insensitive to wether these factors are included or not, as soon as they are non-singular. More precisely, consider the Green's function $1/x^{2k}$ and regularize it first directly as prescribed by \eqref{many-boxes}. Alternatively, one can represent it as,
\begin{equation}
  \frac{1}{x^{2k}}\equiv x^{2p}\frac{1}{x^{2(k+p)}}=-\frac{x^{2p}}{4^{k+p-1}(k+p-1)!(k+p-2)!}
  \Box^{k+p-1}\frac{\ln\mu^2x^2}{x^2}.
\end{equation}

This sort of regularization produces the following contribution to the dilatation operator (upon acting by $\mu\pd/\pd\mu$),
\begin{equation}\label{indirect}
  \sim\frac{\pi^2}{4^{k+p-2}(k+p-1)!(k+p-2)!}x^{2p}\Box^{k+p-2}\delta(x),
\end{equation}
in contrast to
\begin{equation}\label{direct}
  \sim\frac{\pi^2}{4^{k-2}(k-1)!(k-2)!}\Box^{k-2}\delta(x),
\end{equation}
which is given by the direct regularization. The situation is, that both \eqref{indirect} and \eqref{direct} are equivalent when integrated with a regular function.

To prove the equivalence of \eqref{indirect} and \eqref{direct} it is enough to prove the following relation,
\begin{equation}\label{delta-equiv}
  x^{2p}\Box^{k+p-2}\delta(x)=4^{p}\frac{(k+p-1)!(k+p-2)!}{(k-1)!(k-2)!}\Box^{k-2}\delta(x).
\end{equation}
This can be easily done by writing the Fourier transform of the Eq. \eqref{delta-equiv} in spherical coordinates.

\bibliographystyle{hunsrt}
\bibliography{adscft}
\end{document}